\documentclass[conference]{IEEEtran}
\IEEEoverridecommandlockouts
\usepackage{cite}
\usepackage{amsmath,amssymb,amsfonts}
\usepackage{algorithmic}
\usepackage{graphicx}
\usepackage{textcomp}
\usepackage[table,x11names]{xcolor}
\usepackage{bookmark}
\usepackage{microtype}
\usepackage{booktabs}
\usepackage[mode=text]{siunitx}[=2021-04-09]
\usepackage{textgreek}
\usepackage{url}
\usepackage{multirow}
\usepackage{multicol}
\usepackage{tcolorbox}
\usepackage{makecell}
\usepackage{stfloats}
\usepackage{enumitem}

\definecolor{vlightgray}{gray}{0.9}

\newtcolorbox{summarybox}[1][]
{
    sharp corners,
    left=0mm,
    right=0mm,
    boxrule=0.3mm,
    colback=white
    #1,
}

\usepackage{hyperref}
\hypersetup{
 colorlinks=true,   
 citecolor=black,
 urlcolor=blue,
 linkcolor=black,
 bookmarksnumbered=true,     
 bookmarksopen=true,         
 bookmarksopenlevel=1,          
 pdfstartview=Fit,           
 pdfpagemode=UseOutlines,    
 pdfpagelayout=TwoPageRight 
}

\newcolumntype{P}[1]{>{\raggedright\arraybackslash}p{#1}}

\newcommand\hypothesis[2]{\vspace{0.3em} \noindent \hangindent=1em \textbf{Hypothesis #1 (H#1).} {#2}\vspace{0.3em}}

\newcommand{\review}[1]{{\color{black}{#1}}}

\newcommand{\ksedit}[1]{{\color{black}{#1}}}

\def\BibTeX{{\rm B\kern-.05em{\sc i\kern-.025em b}\kern-.08em
    T\kern-.1667em\lower.7ex\hbox{E}\kern-.125emX}}

\begin{document}

\title{Do I Belong? Modeling Sense of Virtual Community Among Linux Kernel Contributors
}




\author{\IEEEauthorblockN{Bianca Trinkenreich}
\IEEEauthorblockA{
\textit{Northern Arizona University}\\
\textit{Flagstaff, AZ, USA} \\
\textit{bianca.trinkenreich@nau.edu}}\\

\IEEEauthorblockN{Daniel M. German}
\IEEEauthorblockA{
\textit{University of Victoria}\\
\textit{Victoria, Canada}\\
\textit{dmg@uvic.ca}}
\and
\IEEEauthorblockN{Klaas-Jan Stol}
\IEEEauthorblockA{\textit{Lero, the SFI Research Centre for Software}\\
\textit{University College Cork, Ireland}\\
\textit{k.stol@ucc.ie}}\\

\IEEEauthorblockN{Marco A. Gerosa}
\IEEEauthorblockA{
\textit{Northern Arizona University}\\
\textit{Flagstaff, AZ, USA}\\
\textit{marco.gerosa@nau.edu}}
\and
\IEEEauthorblockN{Anita Sarma}
\IEEEauthorblockA{
\textit{Oregon State University}\\
\textit{Portland, OR, USA} \\
\textit{anita.sarma@oregonstate.edu}}\\

\IEEEauthorblockN{Igor Steinmacher}
\IEEEauthorblockA{
\textit{Northern Arizona University}\\
\textit{Flagstaff, AZ, USA} \\
\textit{igor.steinmacher@nau.edu}}
}
\vspace{-10mm}

\maketitle

\begin{abstract}
The sense of belonging to a community is a basic human need that impacts an individual's behavior, long-term engagement, and job satisfaction, as revealed by research in disciplines such as psychology, healthcare, and education. Despite much research on how to retain developers in Open Source Software (OSS) projects and other virtual, peer-production communities, there is a paucity of research investigating what might contribute to a sense of belonging in these communities. To that end, we develop a theoretical model that seeks to understand the link between OSS developer motives and a Sense of Virtual Community (SVC). 
We test the model with a dataset collected in the Linux Kernel developer community (N=225), using structural equation modeling techniques.
Our results for this case study show that intrinsic motivations (social or hedonic motives) are positively associated with a sense of virtual community, but living in an authoritative country and being paid to contribute can reduce the sense of virtual community. 
Based on these results, we offer suggestions for open source projects to foster a sense of virtual community, with a view to retaining contributors and improving projects' sustainability. 

\end{abstract}


\section{Introduction}
\label{intro}

The sustainability and long-term survival of Open Source Software (OSS) projects depend not only on attracting but, more crucially, retaining motivated developers \cite{roberts2006understanding}. The reasons behind a developer's decision to stay or leave an OSS project can depend on different intrinsic or extrinsic factors, including an individual's feelings of identity and belonging to the community \cite{barcomb2019episodic}. Hagerty et al. defined a sense of belonging as \textit{``the experience of personal involvement in a system or environment so that persons feel themselves to be an integral part of that system or environment''} \cite{hagerty1992sense}. The need to belong is a powerful, fundamental, and pervasive force that has multiple strong effects on emotional patterns and cognitive processes across all cultures and different types of people \cite{baumeister2017need}. Maslow \cite{maslow1943theory} positioned `belonging' as a basic human need, and Hagerty et al. \cite{hagerty1995developing} posited that a sense of belonging represents a unique mental health concept. A sense of belonging is key to productivity, satisfaction, and engagement \cite{baumeister2017need}, and can help to avoid attrition \cite{allen2019making}. In Science, Technology, Engineering, and Mathematics (STEM), a sense of belonging is strongly related to retention \cite{espinosa2011pipelines}, especially for underrepresented groups \cite{happe2021frustrations}.

The sense of belonging that members have towards others within a certain group is known as a \textit{sense of community} \cite{burroughs1998psychological}. The dimensions of a sense of community include feelings of membership and attachment to a group \cite{blanchard2007developing}, a feeling that members matter to one another and to the group \cite{mcmillan1986sense}. The concept of sense of virtual community (SVC) was developed by observing that virtual communities represent a new form of community, in which social relationships are predominantly forged in cyberspace \cite{koh2003sense}.

Understanding SVC in OSS is relevant as it can influence the vitality and sustainability of a community \cite{blanchard2008testing,tonteri2011antecedents}, and is linked to more satisfied, involved, and committed contributors \cite{kim2020impact}. Individuals who develop a psychological and relational contract with a community are supported by a state of being involved, rather than external factors such as earning something or climbing a career ladder and therefore tend to develop a deeper, reciprocal relationship with that community \cite{burroughs1998psychological}.


Since sustainability is a key concern for OSS projects, we must understand SVC in OSS communities. While several studies have investigated different motivations to contribute to OSS \cite{von2012carrots,harsworking,ghosh2002free,lakhani2003hackers,hertel2003motivation,roberts2006understanding,gerosa2021shifting}, none have modeled how these factors can help or hinder in creating a sense of virtual community. Without a deeper understanding of how the different factors interplay to create a sense of community, strategies that aim to promote individual factors will likely be unsuccessful in creating a sustainable community. Understanding how different factors work together or against each other can help communities strategize how to retain their contributors. Therefore, in this paper, we ask the following research question:

\vspace{1.5mm} \noindent\hangindent1em\textbf{Research Question:} \textit{How does a sense of virtual community develop in Open Source Software projects?}
\vspace{1.5mm}

We answer our research question by first developing a theoretical model of SVC grounded in prior literature (Sec.~\ref{sec:theory_development}). We then evaluate our model through a sample (N=225) of Linux Kernel project contributors, using partial least squares structural equation modeling (PLS-SEM)  (Sec.~\ref{sec:resdesign}). The results of our analysis provide empirical support for part of our model, showing that \textit{hedonism} (motivation that aims to maximize pleasure and fun and minimize pain \cite{tamilmani2019battle}) and \textit{social motives} (motivation that aims to maximize joint gains and others' gains \cite{mcclintock1972social}) have a positive association with a sense of virtual community, which can be weakened when contributors are \textit{being paid} or are surrounded by \ksedit{an}
\review{authoritative culture, i.e., national culture with a high index of power distance (Sec.~\ref{sec:results}).} We conclude the paper by discussing the implications of our findings, and threats to validity (Sec.~\ref{sec:discussion}). 


\section{Background}
\label{sec:background}


\subsection{Sense of Virtual Community}
\label{sec:sense_of_virtual_community}

While numerous definitions of the term `community' exist, a common theme is that it involves human relationships based on some common characteristics \cite{gusfield1975community}. The classical McMillan and Chavis \cite{mcmillan1986sense} definition of `Sense of Community' includes four characteristics: (1) feelings of membership (belonging to, and identifying with, the community), (2) feelings of influence (having an influence on, and being influenced by the community), (3) integration and fulfillment of needs (being supported by others in the community while also supporting them), and (4) shared emotional connection (relationships, shared history, and a `spirit' of the community). \textit{Virtual} communities typify a relatively new form of interaction whereby community members share information and knowledge in the virtual space for mutual learning, collaboration, or problem solving \cite{koh2003sense}. 

The development of OSS involves distributed problem solving within a virtual community \cite{martinez2014current}. Virtual communities are a particularly important type of virtual group because they are self-sustaining social systems in which members engage and connect with each other, developing a Sense of Virtual Community (SVC) \cite{rheingold2000virtual}. The sense of community includes membership, identity, belonging, and attachment to a group that primarily interacts through electronic communication \cite{blanchard2007developing,chang2016mediating,brown2019group}. SVC has been tailored to virtual communities by deriving from McMillan's theory of sense of community \cite{mcmillan1986sense}. The goal of measuring SVC is to assess the ``community-ness'' of a virtual community \cite{blanchard2007developing}.

Community managers can assess and promote SVC to fulfill a core set of members' needs \cite{sutanto2011eliciting}, so they feel they belong to a unique group. Such meaningful relationships are associated with increased satisfaction and communication with the virtual community, trust \cite{blanchard2002sense}, and social capital in the project \cite{zhao2012cultivating}. SVC has been shown to lead to an occupational commitment \cite{blanchard2011sense}, and ultimately can help retain contributors and further attract potential newcomers \cite{blanchard2007developing,chen2013sense}, who are critical to the sustainability of OSS projects \cite{steinmacher2016overcoming}.


\subsection{Motivations to Contribute to Open Source Software}
\label{sec:motivations}

The software engineering literature suggests that, by managing developers' motivation and satisfaction, a software organization can achieve higher productivity levels and avoid turnover, budget overflows, and delivery delays \cite{francca2011motivation}. 
Motivations for joining Open Source has been the topic of considerable research \cite{von2012carrots,harsworking,ghosh2002free,lakhani2003hackers,hertel2003motivation,roberts2006understanding,gerosa2021shifting}; motivations can be extrinsic or intrinsic. 
Extrinsic motivations are based on outside incentives that make people change their actions due to an external intervention \cite{frey1997relationship}. As many companies, including Microsoft, Google, and IBM, hire or sponsor OSS contributors \cite{o2021coproduction}, career ambition and payment have become common extrinsic motivations \cite{schaarschmidt2018company}. However, intrinsic motivations also explain much of contributors' motivations \cite{gerosa2021shifting}, moving a person to act for fun or enjoy a challenge, kinship, altruistic reasons, or ideology, rather than in response to external pressures or rewards \cite{ryan2000self}.  

Previous research showed that several forces influence the decision of an OSS contributor to join, remain, or leave an OSS project \cite{steinmacher2014attracting,kaur2022exploring,calefato2022will}. Despite the extensive attention this topic has received, there are still no studies investigating how OSS contributors driven by different motivations develop a sense of virtual community. We argue that understanding how a sense of virtual community develops in OSS involves understanding the relationship between individual characteristics and motivations and the resulting community-related feelings.

\section{Theory Development}
\label{sec:theory_development}

Feelings of belonging in an online community can be influenced by several individual characteristics and factors of the surrounding environment \cite{allen2020psychology}. In the education literature, researchers \cite{goodenow1993relationship,solomon1996creating} found associations between students' sense of belonging and a range of motivational variables. Motivational factors can be regarded as expectations related to the interaction with a virtual community (answering \textit{why} users behave). Integration and fulfillment of needs refer to the idea that common needs, goals, and beliefs provide an integrative force for a cohesive community that can meet collective and individual needs. Thus, meeting members' needs is a primary function of a strong community \cite{mcmillan1986sense}. 

Individuals who develop a psychological relationship contract with a community because it is focused on a state of being involved---rather than earning something or getting somewhere---tend to develop a sense of community \cite{burroughs1998psychological}. Previous research on online communities also showed that individuals who are driven by \textit{social motives} \cite{neel2016individual} tend to develop a sense of virtual community \cite{kim2016engaging,chang2016mediating}. Based on the Fundamental Social Motives Inventory, we included both kinship and altruism as social motives \cite{neel2016individual} and propose the following hypothesis:

\hangindent1em \hypothesis{1}{Open Source contributors motivated by social \ksedit{reasons} have a higher sense of virtual community}.

Most of the respondents in Gerosa et al.'s study (91\%) agreed (or strongly agreed) that they contribute to OSS for entertainment (fun) \cite{gerosa2021shifting}. Hedonic motivation is a type of motivation that aims to maximize pleasure and fun and minimize pain. It is an umbrella term that includes hedonic expectancy, perceived enjoyment, and playfulness \cite{tamilmani2019battle}. Considering that expectations of enjoyable experiences, feelings of amusement, and being mentally or intellectually stimulated by interactions are associated with a sense of virtual community \cite{koh2003sense,tonteri2011antecedents}, and that changes in the perceived fulfillment of their entertainment needs can determine the change of their sense of virtual community \cite{sutanto2011eliciting}, we propose the following hypothesis:

\hangindent1em \hypothesis{2}{Open Source contributors motivated by hedonic reasons have a higher sense of virtual community}.

It is known that some open source contributors have a strong ideological basis for their actions \cite{stewart2006impact}, believing, for example, that source code should be freely available. Recently, Gerosa et al.'s study showed that, however, ideology is not a popular motivation---especially for young contributors \cite{gerosa2021shifting}.

Historically, the group-based morality of `fighting' a shared dominant opponent incites the sense of virtual community among contributors \cite{mcgowan2001legal}. This feeling was quite common in the 1990s, when big corporations characterized Open Source as `communism' \cite{lea2000} and Linux as a `cancer' \cite{greene2001}. Besides ideology, we include reciprocity in moral motives, as it represents the moral desire of contributors who aim for social justice by giving back to the community \cite{janoff2018model}.

According to the Social Identity theory \cite{tajfel2004social}, sharing a moral vision is positively associated with feelings of belonging. Moreover, a homogeneous ideology throughout a religion was shown as being positively associated with a sense of virtual community \cite{gan2019change}. Hence, we posit that:

\hangindent1em \hypothesis{3}{Open Source contributors motivated by moral reasons have a higher sense of virtual community}.

Motivations may not always be strong enough to sustain an OSS contributor's participation \cite{fang2009understanding}. Motivations may vary for different groups of people, depending on contextual factors. This implies the existence of moderating factors that change the relationship between motivations and a sense of virtual community. Cognitive Evaluation Theory suggests that feelings of autonomy are positively associated with intrinsic motivations and belonging, while tangible rewards negatively affect intrinsic motivating factors \cite{deci1985cognitive}.

We evaluated the role of a feeling of autonomy using the variable of \textit{power distance} from Hofstede's framework of Country Culture \cite{hofstede2001culture} as a proxy; a lower power distance would reflect in higher autonomy. We also evaluated the exposition to tangible rewards using the variable \textit{is paid}.


People in societies exhibiting a large degree of power distance tend to accept a hierarchical order \cite{hofstede2011dimensionalizing}. In high power distance cultures (where a high power differential between individuals is accepted and considered normal), information flows are usually constrained by hierarchy \cite{hofstede2001culture}. As an important cultural value describing the acquiescent acceptance of authority, power distance has received increasing attention in many domains \cite{fock2013moderation,auh2016service}.

Prior research showed that, when surrounded by cultures with a high degree of power distance, students reported a lower sense of belonging to their school \cite{cortina2017school}. Therefore, in hierarchical cultures, leaders need control over the information flow, and the desire to restrict autonomy and access to critical information by lower-level team members could lead to significant organizational barriers to sharing knowledge and working in a community \cite{ardichvili2008learning}. Thus, we define the following moderation hypotheses:

\hypothesis{4a}{Power distance moderates the association between Open Source contributors' social motives and their sense of virtual community}.

\hypothesis{4b}{Power distance moderates the association between Open Source contributors' hedonic motives and their sense of virtual community}.

\hypothesis{4c}{Power distance moderates the association between Open Source contributors' moral motives and their sense of virtual community}.


The traditional notion that OSS developers are all volunteers is now long outdated; many OSS contributors are currently paid, usually employed by a company, to contribute \cite{schaarschmidt2018company,taylor2022love,trinkenreich2020hidden}. Indeed, many Linux Kernel contributors are paid to make their contributions, compensated by firms that have business models relying on the Linux Kernel \cite{corbet2012linux,riehle2014paid,homscheid2015private}. 

In contrast to traditional paid software development work, and despite its benefits to OSS contributors, introducing financial incentives in OSS communities create complex feelings among OSS developers \cite{sharma2022motivation}. For example, developers on the Debian project expressed negative emotion because they felt payment went against the project's espoused values \cite{gerlach2016exploratory}. On the other side, not receiving pay for their work to support their livelihoods can frustrate OSS developers and affect their contributions \cite{sharma2022motivation}.

Despite compensation, OSS contributors may be driven towards a project by both simultaneous feelings of belonging (intrinsic) and payment (extrinsic) \cite{roberts2006understanding,schaarschmidt2018company}. Nevertheless, there is no research examining the complex impact of receiving payment on intrinsic factors associated with SVC. As many OSS developers are currently paid, we would expect that the behavior of those who are paid and those who are not (volunteers) would diverge. Hence, we propose the following three moderating hypotheses:

\hypothesis{5a}{Being paid moderates the association between Open Source contributors' social motives and their sense of virtual community}.

\hypothesis{5b}{Being paid moderates the association between Open Source contributors' hedonic motives and their sense of virtual community}.

\hypothesis{5c}{Being paid moderates the association between Open Source contributors' moral motives and their sense of virtual community}.



\section{Research Design}
\label{sec:resdesign}

\begin{figure}[!b]
\centering
\includegraphics[width=\columnwidth]{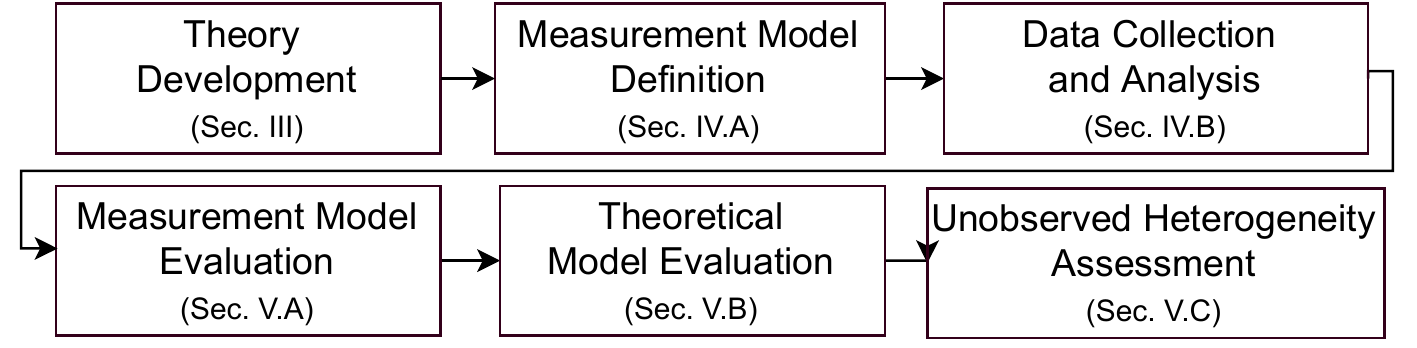}
\caption{Research Design and Phases for Results' Analysis}
\vspace{-3mm}
\label{fig:research_design}
\end{figure}

The research design is summarized in Fig.~\ref{fig:research_design}. We conducted a survey among Linux Kernel contributors to evaluate our theoretical model. We studied one specific community to avoid confounding factors related to differences that each OSS community can pose. Introduced in 1991, the Linux Kernel represents one of the largest and most active OSS projects \cite{homscheid2020linux}, boasting over ten million source lines of code and more than 20,000 contributors from different countries and cultural backgrounds, including volunteers and paid developers from more than 200 companies \cite{tan2020scaling,LinuxKernel2020}. Linux Kernel's is impact is perceived in terms of processes and infrastructure tools that emerged from the community~\cite{LinuxKernel2020}. 
While the Linux Kernel Mailing List is known for its uncivil comments and toxic discussions that tend to discourage people from joining the community \cite{miller2022did}, community leaders aim to change the project's image and increase the sense of community among members. We closely collaborated with contributors and maintainers of the Linux Foundation involved with Linux Kernel, who had a crucial role in designing the data collection instrument and reaching out to potential participants. They engaged in several meetings with the team and reviewed the items of the questionnaires to provide their feedback, making sure that the instrument was appropriate for the study goals. They also distributed the survey to the Linux kernel community, playing an essential role in recruiting the participants for this study.

We used Partial Least Squares--Structural Equation Modeling (PLS-SEM) to analyze the relationships between motivations \cite{ringle2015structural} and a sense of virtual community. SEM is a second-generation multivariate data analysis method; a recent survey (which also provides an introduction to the method) indicates that PLS-SEM has been used to study a variety of phenomena in software engineering \cite{russo2021pls}. SEM facilitates the simultaneous analysis of relationships among constructs, each measured by one or more indicator variables. The main advantage of SEM is being able to measure complex model relationships while accounting for measurement error when using latent variables (e.g., Sense of Virtual Community). PLS-SEM has previously been used in literature to evaluate factors that impact the sense of belonging in other contexts \cite{chen2014building,ellonen2013positive}.

In the following, we discuss the measurement model (i.e., operationalization of constructs), data collection, and analysis.

\subsection{Measurement model}
\label{sec:measurement_model}

The theoretical model comprising the hypotheses is based on a number of theoretical concepts; some of the concepts may be directly observed (e.g., `is paid'), but others cannot (e.g., sense of virtual community)---these concepts are represented as \textit{latent} variables. A latent variable cannot be directly measured or observed but instead is measured through a set of indicators or manifest variables. For the latent variables in this study, we adapted existing measurement instruments.

\textbf{Sense of Virtual Community}: We adapted items about feelings of a virtual community from Blanchard's \cite{blanchard2007developing} instrument of sense of virtual community to better fit with the context of OSS contributions. In collaboration with a group of Linux Kernel community managers, we analyzed the items proposed by Blanchard et al. \cite{blanchard2011sense} and decided to use a subset of questions to compose a shorter version of the instrument to cover the dimensions of SVC. The subset was discussed synchronously by researchers and managers, and the items were considered appropriate and meaningful to represent SVC to the Linux Kernel contributors.

\textbf{Intrinsic Motivations}: We created items based on Gerosa et al. \cite{gerosa2021shifting}'s instrument, which was built upon previous studies of motivations in OSS \cite{lakhani2003hackers,ghosh2002free,harsworking}. Following the community managers' request to make the questionnaire as short as possible, we grouped the intrinsic motivation factors from Gerosa et al.'s study \cite{von2012carrots} into three factors: 1. Social Motives (Kinship and Altruism) \cite{neel2016individual}; 2. Hedonic Motives (Joy and Fun) \cite{tamilmani2019battle}; and 3. Moral motives (Ideology and Reciprocity) \cite{janoff2018model}.

\textbf{English Confidence}: We reused four questions about the self-confidence of fluency levels during interactions by speaking and writing in technical and non-technical situations from a previous survey \cite{steinmacher2021being}.

\textbf{Power Distance:} We asked in which country the respondent lived and used the value per country proposed by Hofstede's framework \cite{hofstede2001culture} as the Power Distance dimension in the model.

For the demographic questions, we adapted questions from surveys used in OSS communities to ask about tenure, self-identified gender, and compensation \cite{bitergia2016survey,GitHubOpenSourceSurvey2017}. \review{Tenure was shown in 10-year slices in Table \ref{tab:demographics} for presentation purposes, but was included as a continuous variable in our analysis. We provided a dropdown list of years since 1991 (the year when the Linux Kernel was launched) for respondents to inform the year they started contributing to the project.}

\subsection{Data Collection and Analysis}
\label{sec:data_collection_and_analysis}

We administered the online questionnaire using LimeSurvey, a leading Open Source survey software, to survey Linux Kernel contributors. We explored their motivations and their sense of virtual community. Our online appendix provides the instrument and replication package \cite{replication_package}.

\subsubsection{Designing the instrument} 
The questions were discussed during 12 online meetings between October 2020 and February 2021 in a group of five researchers experienced in OSS and survey studies and two members of the Linux community. The group discussed each of the questions until reaching a consensus. The questionnaire provides informed consent followed by closed questions about the importance of each motivation factor as a reason to contribute to the Linux Kernel and questions about their feelings about the Linux Kernel community. Finally, we added demographic questions aiming to segment analysis and understand the phenomenon considering the different dimensions of our participants, and an open question for additional comments. Investigating the forces that push people with different individual characteristics can help us better support a diverse community \cite{gerosa2021shifting}. The questions included gender identity, financial compensation, starting the year at Linux Kernel, and country of residence.

\subsubsection{Piloting the questionnaire} 
After designing the instrument, we piloted the questionnaire before distributing it to the population of interest. In the first round, our collaborators from the Linux Foundation recruited a group of Linux Kernel maintainers, who answered the questionnaire and provided feedback.
Although the feedback was overall positive, maintainers suggested reverse-coding some items for the SVC construct, i.e., items worded as negative statements (low score indicates agreement). Inverse, negative, or reverse-coded items can be defined as those having a directionality as opposed to the logic of the construct being measured \cite{weijters2012misresponse}. 
We agreed with the suggestion and inverted two of the four items as this can help to mitigate acquiescence bias \cite{baumgartner2001response}, which can occur when participants tend to agree with statements without regard for their actual content or due to laziness, indifference, or automatic accommodation to a response pattern \cite{podsakoff2003common}. The item \textit{I feel at home in the group} was changed to \textit{I don't feel at home in the group}. We inverted and adapted the question \textit{I feel that my contribution is valued} to \textit{I want to contribute more but I do not feel valued}. 

After the first pilot, we ran two more pilot sessions with two researchers who are Open Source contributors. We used the think-aloud method \cite{van1994think} and recorded their suggestions while answering the questions. We made minor changes to the questionnaire and increased font size for better readability on different devices. 

\subsubsection{Recruiting participants} 
The Linux Foundation contributors who collaborated in this study took the lead in recruiting the participants from the Linux Kernel. They reached out to maintainers and contributors using mailing lists from the Linux Kernel community and interacted with maintainers to ask for engagement. Further, we presented the study motivation during the first day of the Linux Plumbers annual conference ({\url{https://lpc.events/event/11/}}), inviting participants to answer the questionnaire. The survey was available between August 12 and September 21, 2021.

\subsubsection{Sample Analysis}
\label{sec:sample_analysis}
We received 318 responses and carefully filtered the data to consider only valid responses. Respondents who did not complete the whole questionnaire were dropped (n=26). We also dropped the participants who answered ``I'm not sure'' to any of the items included with the five-point Likert scale for Motivations (n=16) and Sense of Virtual Community (n=51). \review{In addition to the 5-point Likert scale, we included a 6th alternative (``I'm not sure") for participants who either preferred not to, or did not know how to, answer the question (to avoid forcing them), which is different from being neutral—based on the dissonance between ignorance and indifference \cite{grichting1994meaning,sturgis2014middle}. Therefore, we considered these responses as missing data. The efficacy of imputation methods has not yet been validated when using FIMIX-PLS; Sarstedt et al. \cite{sarstedt2017treating} recommend removing all responses with missing values for any question before segmenting data into clusters.}

After applying these filters, we retained 225 valid responses from residents of five different continents with a broad tenure distribution. The majority identified as men (84.4\%), from Europe (52.9\%), and paid to contribute (65.4\%), matching previously reported distributions of OSS contributors \cite{GitHubOpenSourceSurvey2017}. Table~\ref{tab:demographics} presents a summary of the demographics.

\begin{table}[!t]
\centering
\caption{Demographics of the Linux Kernel respondents (n=225)}
\label{tab:demographics}
\robustify{\bfseries}
\sisetup{
    mode=text,
    group-digits = false ,
    input-symbols = ( ) [ ] - + *,
    detect-weight=true, 
    detect-family=true,
    table-format=0.2,
    add-decimal-zero=false, 
    add-integer-zero=false,
    round-mode=places, 
    round-precision=1, 
    parse-numbers = true
}
\begin{tabular}{p{5.8cm}
                S[table-format=3.0]
                S[table-format=2.1]}
\toprule
Attribute & {N} & {Percentage}\\
\midrule
\multicolumn{3}{c}{Gender}\\
\midrule
Man & 190 & 84.4\si{\percent}\\
Woman & 21 & 9.4\%\\
Non-binary & 5 & 2.2\%\\
Prefer not to say & 8 & 3.6\%\\
Prefer to self describe & 1 & 0.4\%\\

\midrule
\multicolumn{3}{c}{Continent of Residence}\\
\midrule
Europe & 119 & 52.9\%\\
North America & 68 & 30.2\%\\
Asia & 32 & 14.2\%\\
South America & 6 & 2.7\%\\
\midrule
\multicolumn{3}{c}{Starting year at the Linux Kernel}\\
\midrule
2000 or earlier & 28 & 12.4\%\\
Between 2001 and 2010 & 77 & 34.2\%\\
Between 2011 and 2021 & 120 & 53.4\%\\
\midrule
\multicolumn{3}{c}{Current Compensation for the Linux Kernel contributions}\\
\midrule
Paid & 145 & 64.4\%\\
Unpaid (volunteer) & 80 & 35.6\%\\
\bottomrule
\end{tabular}
\vspace{-5mm}
\end{table}

To establish an appropriate sample size, we conducted a power analysis using the free G*Power tool \cite{faul2009statistical} (see online appendix for details). 
The maximum number of predictors in our model is six (three motivations and three control variables to SVC). This calculation indicated a minimum sample size of 62 and our sample of 225 exceeded that number considerably.

We used the software package SmartPLS version 4 for the analyses. The analysis procedures for PLS-SEM comprise two main steps, with tests and procedures in each step. The first step is to evaluate the measurement model, which empirically assesses the relationships between the constructs and indicators (see Sec.~\ref{sec:results_measurement_model}). The second step is to evaluate the theoretical (or structural) model that represents the hypotheses (see Sec.~\ref{sec:results_structural_model}).

\section{Analysis and Results}
\label{sec:results}

In this section, we describe our results, which include the evaluation of the measurement model (Sec.~\ref{sec:results_measurement_model}), followed by the hypotheses evaluation in the structural model (Sec.~\ref{sec:results_structural_model}), both computed through our survey data. We assess the significance of our model by following the evaluation protocol proposed by previous research \cite{hair2019use,russo2021pls} to make results consistent with our claims. The path weighting scheme was estimated using SmartPLS 4 \cite{sarstedt2019partial}.


Two tests are recommended to ensure that a dataset is suitable for factor analysis \cite{bartlett1950tests,hair1995multivariate}.
We first conducted Bartlett's test of sphericity \cite{bartlett1950tests} on all constructs. We found a p-value \textless{} .01 (p values less than .05 indicate that factor analysis may be useful). Second, we calculated the Kaiser-Meyer-Olkin (KMO) measure of sampling adequacy. Our result (.81) is well above the recommended threshold of .60 \cite{hair1995multivariate}.

\subsection{Evaluation of the Measurement Model}
\label{sec:results_measurement_model}

\begin{figure*}[!b]
\centering
\includegraphics[width=0.95\textwidth]{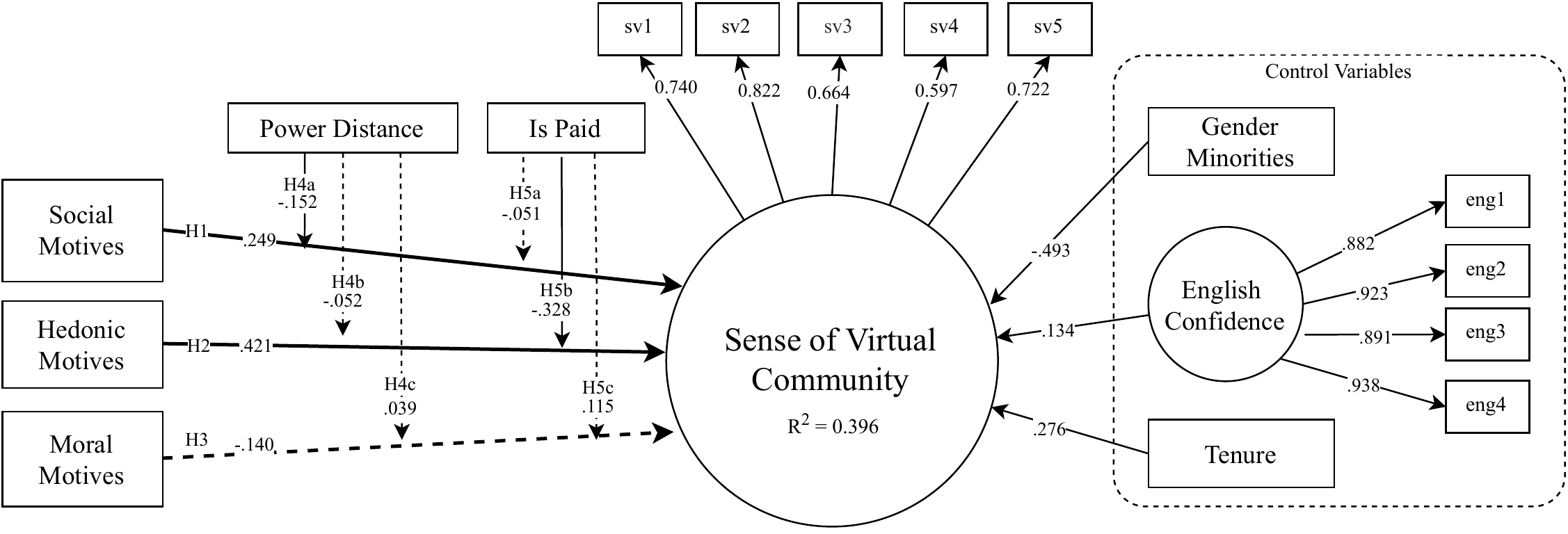}
\caption{Item loadings and path coefficients (p $<$ 0.05 indicated by a full line). Non-significant links are indicated with a dashed line}
\label{fig:evaluating_structutal_model}
\end{figure*}

Some of the constructs in the theoretical model (see Fig.~\ref{fig:evaluating_structutal_model}) are modeled as latent variables, i.e., measured by more than one observed variable (i.e., item/question on the survey). The first step in evaluating a structural equation model is to assess the soundness of the measurement of these latent variables---this is referred to as evaluating the `measurement model' \cite{hair2019use}. We present the assessment of several criteria.

\subsubsection{Convergent Validity}

First, we assess whether the questions (indicators) that represent each latent variable are understood by the respondents in the same way as they were intended by the designers of the questions \cite{kock2014advanced}, i.e., we assess the convergent validity of the measurement instrument. The assessment of convergent validity relates to the degree to which a measure correlates positively with alternative measures of the same construct. Our model contains two latent variables, both of which are reflective (not formative), as functions of the latent construct. Changes in the theoretical, latent construct are reflected in changes in the indicator variables \cite{hair2019use}.

We used two metrics to assess convergent validity: the Average Variance Extracted (AVE) and the loading of an indicator onto its construct (the outer loading).

The AVE is equivalent to a construct's communality \cite{hair2019use}, which is the proportion of variance that is shared across indicators. A reflective construct is assumed to reflect (or ``cause'') any change in its indicators. The AVE should be at least .50, indicating that it explains most of the variation (i.e. 50\% or more) in its indicators \cite{hair2019use}. This variance is indicated by taking the squared value of an indicator's loading. 
\review{As Table~\ref{tab:internal_consistency_reliability} shows,} all AVE values for both latent constructs in our model are above this threshold of .50.

\begin{table}[!hb]
\centering
\caption{Internal Consistency Reliability}
\label{tab:internal_consistency_reliability}
\robustify{\bfseries}
\sisetup{
    mode=text,
    group-digits = false ,
    input-signs ={-},
    input-symbols = ( ) [ ] - + *,
    detect-weight=true, 
    detect-family=true,
    table-format=0.2,
    add-decimal-zero=false, 
    add-integer-zero=false,
    round-mode=places, 
    round-precision=3, 
    parse-numbers = true
}
\begin{tabular}{p{3.2cm}
                S
                S
                S}
\toprule
& {\makecell{Cronbach's \textalpha}} & {\makecell{CR}} & {AVE}\\
\midrule
English Confidence & .932 & .951 & .830\\
Sense of Virtual Community & .767 & .839 & .511\\
\bottomrule

\end{tabular}
\end{table}

A latent variable is measured by two or more indicators; indicators with loading below .4 should be removed because this implies that a change in the latent construct that it purportedly represents (or `reflects') does not get reflected in a sufficiently large change in the indicator \cite{hair2019use}. Outer loading of .7 is widely considered sufficient, and .6 is considered sufficient for exploratory studies \cite{hair2019use}. We followed an iterative process to evaluate the outer loading of the latent constructs; the indicators of the construct English Confidence all exceeded .7, but SVC had two indicators below .7. We removed the SVC indicator which had a loading below .4 (\textit{svc6: a majority of the developers and I want the same thing}). After removing this indicator, the AVE value of SVC (now with five indicators) increased from .44 to .51 and all outer loadings were above .60.

\subsubsection{Internal Consistency Reliability}

Second, we verified how well the different indicators are consistent with one another and can reliably and consistently measure the constructs, i.e., we assess the internal consistency reliability. A high degree of consistency means that the indicators refer to the same construct. There are several tests to measure internal consistency reliability. We performed both the Cronbach's \textalpha{} and Composite Reliability tests; Cronbach's \textalpha{} frequently shows lower values, whereas the Composite Reliability (CR) is a more liberal test, which sometimes overestimates the values \cite{hair2019use}.
A desirable range of values for both Cronbach’s \textalpha{} and CR is 
between .7 and .9 
\cite{hair2019use}. Values below .6 suggest a lack of internal consistency reliability, whereas values over .95 suggest that indicators are too similar and thus not desirable. 
The Cronbach \textalpha{} and CR values for both latent variables fell in the range .75--.95; only the CR for English Confidence was slightly over at .951. AVE values were both higher than .50.


\subsubsection{Discriminant Validity}

Third, we verified whether each construct represents characteristics not measured by other constructs, i.e., we assessed the discriminant validity of the instrument (indicating the distinctiveness of the constructs). Our model includes two latent variables (SVC and English Confidence). A primary means to assess discriminant validity is to investigate the Heterotrait-monotrait (HTMT) ratio of correlations, developed by Henseler et al. \cite{henseler2015new}. The discriminant validity could be considered problematic if the HTMT ratio exceeds .9 \cite{henseler2015new}; some scholars recommend a more conservative cut-off of .85 \cite{hair2019use}. The HTMT ratio between the two latent constructs (SVC and English Confidence) was .24. We also assessed the cross-loadings of indicators and the Fornell-Larcker criterion. \review{Items should only load onto their `native' construct, the one they purportedly represent (Table \ref{tab:cross_loadings}).}

\begin{table}[!h]
\centering
\caption{Cross loadings of the retained indicators on the constructs}
\label{tab:cross_loadings}
\begin{tabular}{p{0.3cm}
                P{6.2cm}
                >{\centering\arraybackslash}P{0.3cm}
                >{\centering\arraybackslash}p{0.3cm}}
\toprule
& & {SVC} & Engl. Conf.\\
\midrule
svc1 & I don't feel at home in the group & \textbf{.740} & .191\\
svc2 & I feel that I belong to the group & \textbf{.822} & .231\\
svc3 & \hangindent1em If I have a problem, I know members in the group who I can ask for
help & \textbf{.662} & .063\\
svc4 & I want to contribute more but I do not feel valued & \textbf{.583} & .113\\
svc5 & A majority of the developers in the group know me & \textbf{.722} & .242\\
eng1 & \hangindent1em Participating in a non-technical discussion on the email list & .180 & \textbf{.881} \\
eng2 & Performing Reviews & .216 & \textbf{.925}\\
eng3 & Speaking with others (face to face) & .226 & \textbf{.890}\\
eng4 & Participating in technical discussions on the email list & .278  & \textbf{.939} \\

\bottomrule


\end{tabular}
\end{table}

For the sake of completeness, we report the Fornell-Larcker procedure in the online appendix \cite{replication_package}.


\subsection{Evaluation of the Theoretical Model}
\label{sec:results_structural_model}

We now evaluate and discuss the theoretical model, which includes the evaluation of the hypotheses.

\subsubsection{Assessing Collinearity}

Our theoretical model has three different exogenous variables of intrinsic motivations, the moderators `Compensation' and `Power Distance,' and the control variables `English Confidence,' `Gender,' and `Tenure.' We hypothesized that the exogenous variables are associated with the endogenous variable Sense of Virtual Community. To ensure that the three exogenous constructs are independent, we calculate their collinearity using the Variance Inflation Factor (VIF). A widely accepted cut-off value for the VIF is 5 \cite{hair2019use}, and in our model, all VIF values are below 5.

\subsubsection{Path Coefficients and Significance}

PLS does not make strong assumptions about the distribution (such as a Normal distribution) of the data, so parametric tests of significance should not be used. To evaluate whether path coefficients are statistically significant, PLS packages employ a bootstrapping procedure. This involves drawing a large number (usually five thousand) of random subsamples with replacement. The replacement is needed to guarantee that all subsamples have the same number of observations as the original data set. The PLS path model is estimated for each subsample. 
From the resulting bootstrap distribution, a standard error can be determined \cite{hair2019use}, which can subsequently be used to make statistical inferences. The mean path coefficient determined by bootstrapping can differ slightly from the path coefficient calculated directly from the sample; this variability is captured in the standard error of the sampling distribution of the mean. 

Table~\ref{tab:new_path_analysis} shows the results for our hypotheses, including the mean of the bootstrap distribution (\textit{B}), the standard deviation (\textit{SD}), the 95\% confidence interval, and the p-values. 

\review{The path coefficients in Fig.~\ref{fig:evaluating_structutal_model} and Table \ref{tab:new_path_analysis} are interpreted as standardized regression coefficients, indicating the direct effect of a variable on another. Each hypothesis is represented by an arrow in the diagram in Fig.~\ref{fig:evaluating_structutal_model}. For example, the arrow pointing from Hedonic Motives to SVC represents H2. Given its positive path coefficient (0.421), Hedonic Motives are positively associated with SVC. The path coefficient is 0.421; this means that when the score for ``Hedonic'' motives increases by one standard deviation unit, the score for ``Sense of Virtual Community''  increases by 0.421 standard deviation unit (the standard deviation is the amount of variation of a set of values).}

Based on these results, we found support for Hypotheses H1 (p=.002), H2 (p=.000), H4a (p=.045), and H5b (p=.023). Hypothesis H3 was not supported, nor were H4b, H4c, H5a, or H5b (all p values $>$ .2). The three control variables all have significant associations with SVC: English confidence, gender, and tenure (p \textless{} .05).

\begin{table}[!t]
\centering
\caption{Standarized path coefficients, standard deviations, confidence intervals, and p values}
\label{tab:new_path_analysis}
\robustify{\bfseries}
\sisetup{
    mode=text,
    group-digits = false ,
    input-signs ={-},
    input-symbols = ( ) [ ] - + *,
    detect-weight=true, 
    detect-family=true,
    table-format=0.2,
    add-decimal-zero=false, 
    add-integer-zero=false,
    round-mode=places, 
    round-precision=2, 
    parse-numbers = true
}
\begin{tabular}{P{4.3cm}
                S
                S[table-format=0.2]
                >{\centering\arraybackslash}p{1.3cm}
                S[table-format=0.3,round-precision=3]}
\toprule
& {\textit{B}} & {SD} & {95\% CI} & {\textit{p}}\\
\midrule
H1 Social motives$\rightarrow$SVC & .249 & .105 & (.04, .46) & \bfseries .002 \\
H2 Hedonic motives$\rightarrow$SVC & .421 & .114 & (.19, .64) & \bfseries .000\\
H3 Moral motives$\rightarrow$SVC & -.140 & .112 & (-.36, .08) & .215\\
\midrule
\hangindent1em H4a Power distance \texttimes{} social motives $\rightarrow$ SVC & -.152 & .076 & (\num{-.31},\num{-.01}) &  \bfseries .045\\
\hangindent1em H4b Power distance \texttimes{} hedonic motives $\rightarrow$SVC  & -.052 & .074  & (\num{-.18}, .11)  & .477 \\
\hangindent1em H4c Power distance \texttimes{} moral motives$\rightarrow$SVC & .042 & .069  & (\num{-.10}, .17)  & .539\\
H5a is Paid \texttimes{} social motives$\rightarrow$SVC & -.051 & .065  & (\num{-.32}, .20) & .696  \\ 
H5b is Paid \texttimes{} hedonic motives$\rightarrow$SVC & -.328 & .144 & (\num{-.62}, .05)  & \bfseries .023  \\ 
H5c is Paid \texttimes{} moral motives$\rightarrow$SVC & .115 & .137  & (\num{-.17}, .36) & .404 \\ 
\midrule
Gender minorities$\rightarrow$SVC & -.493 & .170  & (\num{-.81},\num{-.14}) & \bfseries .004 \\ 
English confidence$\rightarrow$SVC & .134 & .006 & (.01,.25) & \bfseries .025 \\ 

\bottomrule

\end{tabular}
\vspace{-3mm}\end{table}

\subsubsection{Coefficient of Determination}

We assessed the relationship between constructs and the predictive capabilities of the model. The \textit{R}\textsuperscript{2} values of the endogenous variable in our model (SVC) was 0.4, which is considered weak-moderate \cite{hair2019use,henseler2009use}.

We also inspected Stone-Geisser’s \textit{Q}\textsuperscript{2} 
\cite{stone1974cross} value, which is a measure of external validity, as an indicator of the model’s predictive relevance \cite{hair2019use}, and can be obtained through a so-called blindfolding procedure (available within the SmartPLS software). Blindfolding is a resampling technique that omits certain data, predicts the omitted data points, then uses the prediction error to cross-validate the model estimates \cite{tenenhaus2005pls}. \textit{Q}\textsuperscript{2} values are calculated only for the SVC, the reflective endogenous construct of our model, with a value of .17. Values larger than 0 indicate the construct has predictive relevance, while negative values show the model does not perform better than the simple average of the endogenous variable would do.

\review{The Standardized Root Mean Square Residual (SRMR) is a common fit measure that is appropriate to detect misspecification of PLS-SEM models \cite{russo2021pls}. SRMR is the square root of the sum of the squared differences between the model-implied and the empirical correlation matrix, or the Euclidean distance between the two matrices \cite{henseler2014common}. A value of 0 for SRMR would indicate a perfect fit, and a cut-off value of 0.08 is considered adequate \cite{henseler2016using}. Our results suggest a good fit of the empirical data in the theoretical model (SRMR = 0.06).}

\subsubsection{Moderating Factors}
\label{sec:mod_factors}

We examined our data to determine if the impact of each intrinsic motivation on a sense of virtual community would change when they are exposed to a high Power Distance culture or when they are financially compensated to contribute. 

Only significant results at 0.05 are reported, with confidence intervals calculated through bootstrapping.
\begin{itemize}
 \item Power Distance Country Culture: Being surrounded by a high power distance culture, in which leaders impose a high level of control and restrict the information flow \cite{hofstede2001culture}, has been reported to negatively affect the sense of virtual community \cite{ardichvili2008learning}. We did not find significant correlations between Power Distance and SVC for hedonic or moral motivations. Still, we found it for social motivations, which has a moderating effect on our model. Hence, we found support for H4a but do not support H4b and H4c.
 \item Compensation: Being paid to contribute reduces the sense of virtual community of contributors driven by hedonic motivations but not by social motivations and neither by moral motivations. Hence, we found support for H5b but rejected H5a and H5c.
\end{itemize}

\begin{figure}[!b]
\centering
\includegraphics[trim={0 0 10mm 0},
                clip, width=\columnwidth]{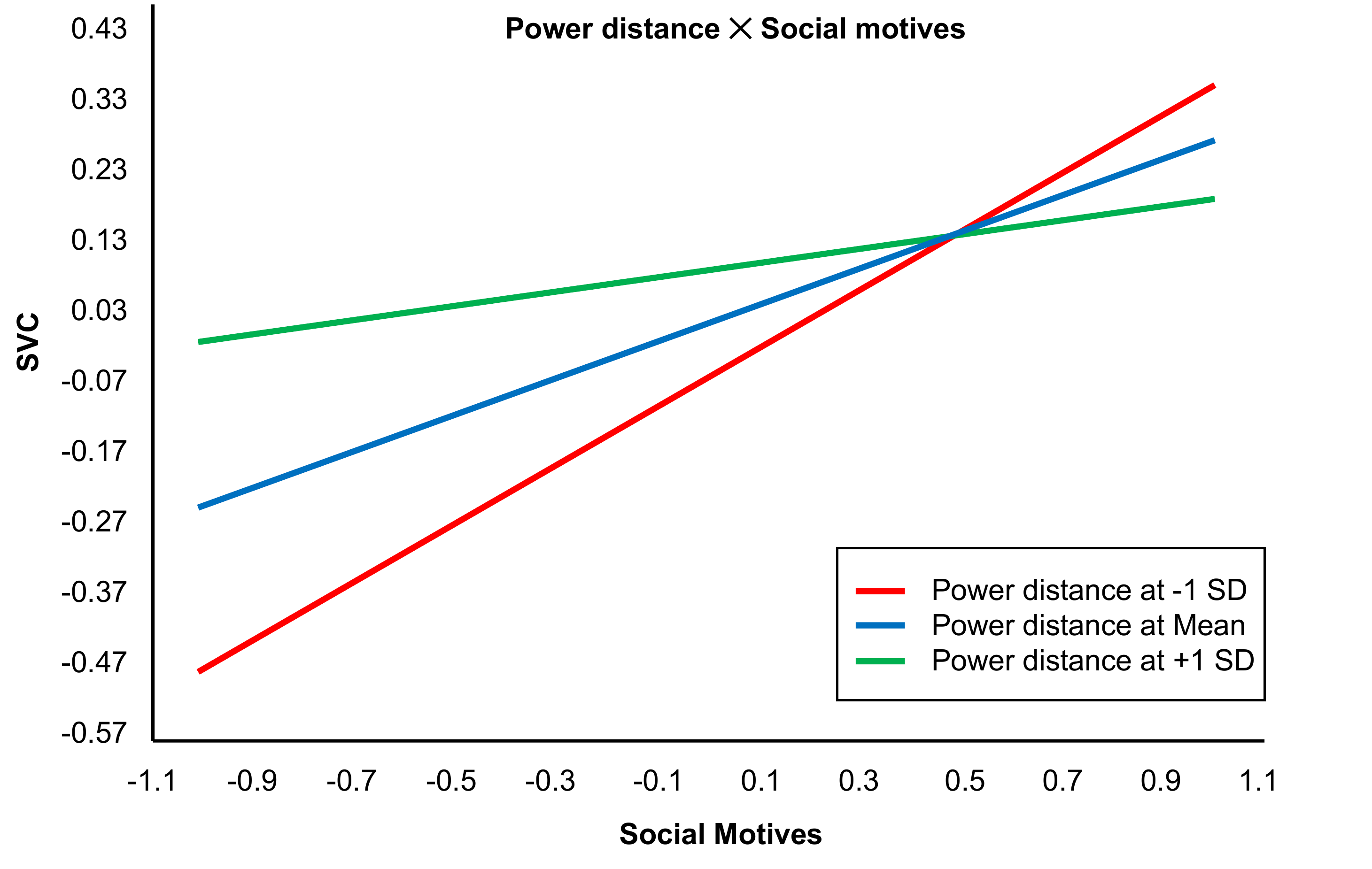}
\caption{Power distance as a moderator of Social motives $\rightarrow{}$ SVC}
\label{fig:slope}
\end{figure}

Fig.~\ref{fig:slope} presents an interaction diagram showing the simple slopes for the relationship between the exogenous variable Social Motives and the endogenous variable SVC. All three slopes are positive, indicating a positive relationship; the top line (in green) is at +1 standard deviation of the moderator, Power distance; the bottom slope (in red) is at \num{-1} standard deviation of the moderator. The middle slope (in blue) represents the relationship at the mean level of Power distance. The figure shows that given a higher level of Power distance, the relationship between social motives and SVC is \textit{dampened} (flatter), whereas with lower levels of Power distance, the relationship is \textit{strengthened} (steeper).


\subsubsection{Control Variables}

We also examined our data to determine if being part of gender minorities, tenure, or English Confidence could strengthen or weaken the sense of virtual community. We found that participants who identify as gender minorities tend to have a lower sense of virtual community, while participants with higher tenure and English Confidence reported a higher sense of virtual community.



\subsection{Cluster Analysis: Detecting Unobserved Heterogeneity}
\label{sec:clusteranalysis}

While moderators and context factors capture \textit{observed} heterogeneity (see Sec.~\ref{sec:mod_factors}), there may also be \textit{unobserved heterogeneity}, or \textit{latent classes} of respondents, the presence of which could threaten the validity of results and conclusions \cite{sarstedt2017treating}. Latent classes of respondents refer to some groupings of respondents on one or more criteria that were not measured. The hypothesis results may differ for different groups. 

We adopted Becker et al.'s approach \cite{becker2013discovering}, which jointly applies PLS-POS and FIMIX algorithms to identify latent classes. In Step~1, we used the minimum sample size for the maximum number of segments and ran FIMIX to find the optimal number of segments. In Step~2, we ran PLS-POS to compute the segmentation. In Step~3, we ran a multi-group analysis (PLS-MGA) and evaluated whether the segments were distinguishable. In Step~4, we checked if the resulting groups were plausible. We discuss the steps in more detail.

In Step~1, we assessed the maximum number of segments according to the minimum sample size (see Sec.~\ref{sec:sample_analysis}). Dividing the sample size (225) by the minimum sample size (62) yields a theoretical upper bound of three segments; each segment should satisfy the minimum sample size. We ran FIMIX for one (meaning, treating the original sample as a single segment), two, and three segments  \cite{sarstedt2017treating}. The results were compared using several different retention criteria (see Table~\ref{tab:fimix}) \cite{sarstedt2017treating}. For each criterion, the optimal solution is the number of segments with the lowest value (in \textit{italics} in Table~\ref{tab:fimix}), except in terms of criterion `EN,' where higher values indicate a better separation of segments. 
Sarstedt et al. \cite{sarstedt2011uncovering} argue that researchers should start the fit analysis by jointly considering the \textit{combination} of modified Akaike's Information Criterion with factor 3 (AIC3) and Consistent AIC (CAIC) (Group 1 in Table~\ref{tab:fimix}): when \textit{both} criteria suggest the same number of segments, this result is likely to be most appropriate. As this is not the case here (AIC3 suggests 3 segments, CAIC suggests 1 segment), a second evaluation considers whether modified Akaike's Information Criterion with factor 4 (AIC4) and Bayesian Information Criterion (BIC) suggest the same number of segments (Group 2 in Table~\ref{tab:fimix}). Again, this is not the case as AIC4 suggests 3 segments, and BIC suggests 1 segment). 
The third evaluation (Group 3) considers the joint analysis of Akaike's Information Criterion (AIC) and Minimum Description Length with factor 5 (MDL5); first, consider the number of segments indicated by the lowest values of AIC (3 segments) and MDL5 (1 segment). The appropriate number of segments should be lower than suggested by AIC (because it tends to overestimate) and higher than the number of segments suggested by MDL5 (because it tends to underestimate). Hence, this combination suggests that a 2-segment solution is appropriate because 2 is lower than the 3 suggested by AIC and higher than the 1 suggested by MLD5. The value of EN is highest for the 2-segment solution.

\begin{table}[!tb]
\centering
\caption{Establishing adequate number of segments}
\label{tab:fimix}

\robustify{\bfseries}
\robustify{\itshape}
\sisetup{
    mode=text,
    group-digits = false ,
    detect-all=true,
    input-signs ={-},
    input-symbols = ( ) [ ] - + *,
    detect-weight=true, 
    detect-family=true,
    table-format=3.3,
    add-decimal-zero=false, 
    add-integer-zero=false,
    round-mode=places,
    round-precision=3, 
    parse-numbers = true
}

\begin{tabular}{p{0.8cm}p{2cm}SSS[table-format=4.3]}
\toprule
Group & Criterion & {1-Segment} & {\bfseries 2-Segment} & {3-Segment}\\
\midrule
\multirow{2}{=}{1}& AIC3 & 574.153 & 540.241 & \itshape 517.508 \\
& CAIC & \itshape 625.395 & 646.141 & 678.065 \\ 
\midrule 
\multirow{2}{=}{2}& AIC4 & 589.153 & 571.241 & \itshape 564.508 \\
& BIC & \itshape 610.395 & 615.141 & 631.065 \\
\midrule 
\rowcolor{vlightgray} 
\cellcolor[gray]{.9} & \bfseries AIC & 559.153 & 509.241 & \bfseries \itshape 470.508 \\
\rowcolor{vlightgray} 
\multirow{-2}{=}{\cellcolor[gray]{.9}3}& \bfseries MDL5 & \bfseries \itshape 935.361 & 1286.737 & 1649.292 \\
\midrule 
4 & EN & 0 & 0.869 & 0.821 \\
\bottomrule
\end{tabular}
\vspace{-5mm}
\end{table}

In Step~2, we evaluated the segment sizes of the 2-segment solution and proportions of data to check whether groups were substantial or candidates for exclusion. A segment is not substantial if its size is considerably lower in proportion (e.g., a 2\% segment size) or below the minimum sample size \cite{becker2013discovering}. The 2-segment solution divided the dataset into groups with 158 (70.2\%) and 67 (29.8\%) observations; both considerable portions and larger than the minimum sample size \cite{becker2013discovering}.


In Step 3, we ran a multi-group analysis (PLS-MGA) with parametric tests to verify whether the segments were distinguishable \cite{becker2013discovering}, i.e., whether the results were different for the two segments. We found significant differences in hypotheses H4b-c, H5a-c, and on the control variables Tenure and English Confidence (see Table~\ref{tab:pls-pos}), thus, we conclude these two segments represent two different groups of respondents. Both groups presented \textit{R}\textsuperscript{2}, goodness-of-fit (GoF), and SRMR \cite{sarstedt2017treating} equal or more favorable than the original model. The values of the path coefficients and the explained variance of the endogenous variable SVC are shown in Table~\ref{tab:pls-pos}, which presents the results for the two segments, as well as the original estimates (see column \textit{B} in Table~\ref{tab:new_path_analysis}).

\begin{table}[b]
\centering
\vspace{-2mm}
\caption{Group Paths Coefficients: coeficients in bold are significant; 
lines in gray show significant difference between segments}
\label{tab:pls-pos}
\robustify{\bfseries}
\sisetup{
    mode=text,
    group-digits = false ,
    input-signs ={-},
    input-symbols = ( ) [ ] - + *,
    detect-weight=true, 
    detect-family=true,
    table-format=0.2,
    add-decimal-zero=false, 
    add-integer-zero=false,
    round-mode=places, 
    round-precision=2, 
    parse-numbers = true
}
\begin{tabular}{P{5cm}
                S[table-format=0.2]
                S[table-format=0.2]
                S[table-format=0.2]}

\toprule

& \multicolumn{2}{c}{2-segment solution} & {Orig.}\\
\cmidrule(l){2-3}
& {Seg. 1} & {Seg. 2} & {All}\\
& {Hedonic} & {Social} & {}\\
\midrule
Sample size (N) & {158} & {67} & {225}\\
Coefficient of determination (\textit{R}\textsuperscript{2}) & .57 & .94 & .40\\
\midrule 
\hangindent1em H1 Social motives $\rightarrow$ SVC & -.04 &  \bfseries.22 & \bfseries .25\\ 
\hangindent1em H2 Hedonic motives$\rightarrow$SVC & \bfseries.31 & .06 & \bfseries .42\\ 
\hangindent1em H3 Moral motives$\rightarrow$SVC & -.03 &  \bfseries-.23 & -.14\\ 

\hangindent1em H4a Power distance \texttimes{} social mot.$\rightarrow$ SVC & -.10 & \bfseries-.24 & \bfseries -.15\\ 
\rowcolor{vlightgray}\hangindent1em H4b Power distance \texttimes{} hedonic mot.$\rightarrow$SVC & \bfseries-.07 & .14 & -.05\\ 
\rowcolor{vlightgray}\hangindent1em H4c Power distance \texttimes{} moral motives$\rightarrow$SVC & -.02 & .22 & .04\\ 
\rowcolor{vlightgray}\hangindent1em H5a is Paid \texttimes{} social motives$\rightarrow$SVC & \bfseries.49 & \bfseries-.61 & -.05 \\ 
\rowcolor{vlightgray}\hangindent1em H5b is Paid \texttimes{} hedonic motives$\rightarrow$SVC & \bfseries -.50 & \bfseries .50 & \bfseries -.33\\ 
\rowcolor{vlightgray}\hangindent1em H5c is Paid \texttimes{} moral motives$\rightarrow$SVC & \bfseries-.15 & .32 & .12 \\ 
\midrule 
\hangindent1em  Gender minorities$\rightarrow$SVC & \bfseries-.70 & \bfseries -.92  & \bfseries -.49\\ 
\rowcolor{vlightgray} \hangindent1em  English confidence$\rightarrow$SVC & -.15 & \bfseries .88 & \bfseries .13\\ 
\rowcolor{vlightgray} Tenure$\rightarrow$SVC & \bfseries .43 & -.05 & \bfseries .28 \\ 

\bottomrule

\end{tabular}
\vspace{-3mm}
\end{table}

In Step 4, we examined that groups were ``plausible'' \cite{becker2013discovering} by explaining the different segments (highlighted in gray in Table~\ref{tab:pls-pos}) to label the segments. This labeling is somewhat speculative and not definitive, not dissimilar to the labeling of emergent factors in exploratory factor analysis. Given that for Segment 1 only Hedonic motives are significant, we posit that this segment represents \textit{Hedonists} (\textit{B}=.31); for Segment 2, we find that Social motives are significant (\textit{B}=.22), thus we label Segment 2 as \textit{Socially Motivated}. We note that moral motives were not significant in the original analysis (see column `Orig.'), but this did become significant with a negative coefficient (\textit{B}=\num{-.23}) for Segment 2. For hedonists (Seg. 1), tenure (\textit{B}=.43) is positively associated with SVC. When social motives are associated with SVC (Seg. 2), English Confidence positively affects SVC (\textit{B}=.88). Both hedonists (\textit{B}=\num{-.50}) and socially motivated (\textit{B}=\num{-.61}) contributors have the association with SVC weakened when they are paid. Both groups showed that being part of a gender minority is associated with less SVC.



\section{Discussion}
\label{sec:discussion}

In this study, we developed a theoretical model grounded in psychology literature to map the relationship between a Sense of Virtual Community and intrinsic motivations in OSS. The theoretical model includes a number of salient factors that have been shown to be important in belonging to an online community in general but not yet within the OSS domain. Over the past two decades, the nature of OSS communities (as a specific type of online community) has changed; traditionally, OSS was male-dominated and primarily volunteer-based, but being paid to contribute is now common, and increasingly we observe the participation of what we refer to as ``minorities'' in the broadest sense of the word, including women \cite{trinkenreich2022women}. Our analysis highlights a number of key findings and implications; as we discuss these quantitative results, we bring exemplar quotes from the respondents' responses to the final open question of the survey to illustrate the discussion.

\textbf{H1. Social motives $\rightarrow$ SVC:} Social motives have a positive association with SVC. The intrinsic social motivations of kinship and altruism are positively associated with a sense of virtual community in OSS. This finding was corroborated by one of our respondents in the final open question, who associated SVC with social motivations: \textit{``I did not fit in, in a big way. I was never able to create enough social capital to make networking effective, no matter who I tried to connect with.''} 
Another respondent mentioned \textit{``not being able to relate to colleagues} and named their perceived lack of SVC as \textit{``a sense of otherness that never goes away.''}
However, the cluster analysis (Sec.~\ref{sec:clusteranalysis}) indicated Segment 1 (which we labeled `hedonic') to be non-significant, but Segment 2 (labeled `social') is significant. We also found that for the `socially motivated' English confidence is much more strongly related (\textit{B}=.88 instead of .13) to SVC. This is intuitive because socially motivated people seek interaction, and English is the primary language within the Linux Kernel community.

\textbf{H2. Hedonic Motives $\rightarrow$ SVC:} Hedonic motives have a positive association with a Sense of Virtual Community. OSS communities should seek to prevent toxic and other types of undesirable behavior that might reduce contributors' enjoyment; communities could also consider setting more clear community codes of conduct
\cite{raman2020stress,cohen2021contextualizing,miller2022did}. The cluster analysis showed that when only hedonism (not social motives) is associated with SVC (Seg. 1), Tenure is also associated with SVC. Hedonic-motivated contributors from our sample are also the ones who have longer tenure associated with SVC. Those contributors may have surpassed the initial barriers \cite{steinmacher2014barriers} and find enjoyment, or as mentioned by another respondent: \textit{``It is therapeutic. When I feel bad about myself, [..] it calms me down emotionally to do Kernel development when I feel like that.''}

\textbf{H3. Moral Motives $\rightarrow$ SVC:} The cluster analysis does not support H3. While social motives are positively associated with SVC (Seg. 2), moral motives are negatively associated with and reduce SVC. The first association is expected and not surprising \cite{neel2016individual}. People motivated by kinship or because they are happy to help others are keener to be part of the team and feel good in a community \cite{kim2016engaging,chang2016mediating}. Interestingly, the SVC presented a negative association with moral motivation. We argue that people motivated by ideological reasons may contribute regardless of how they feel about belonging there. They do it because they feel it is the right thing to do, either because it is the most ethical choice, as advocated by the Free Software Foundation (\url{https://www.fsf.org/}), or because they have a moral debt to the software project that they use, so they pay back \cite{janoff2018model}. Future research can investigate how strong the ties between these people and the community are and what roles they play in building SVC for the rest of the community.  

\textbf{H4a/b. Power Distance moderates the relationship between (a) Social and (b) Hedonic motives to SVC:} Being surrounded by a culture with a high level of power distance weakens SVC for socially motivated contributors (when we consider all contributors). Still, if we consider Segment 1 (Hedonic) in the cluster analysis, we observe that power distance also weakens the SVC associated with hedonism. These results align with Cognitive Evaluation Theory \cite{deci1985cognitive}; contributors driven by hedonic (Seg. 1) or social motives (Seg. 2) need more autonomy (through less hierarchy---less Power Distance) to develop a Sense of Virtual Community. When not exerted in toxic and harsh ways to discipline community members, concerted control of communications can also ultimately play a pro-social role in increasing the SVC by increasing cohesiveness, commitment, and conformity \cite{gibbs2019investigating}.



\textbf{H5a/b. Payment moderates the relationship between (a) Social and (b) Hedonic motives to SVC:} Being paid to contribute weakens the association with SVC for hedonist contributors. The cluster analysis shows that being paid to contribute also weakens the SVC associated with social motives. Paid contributors, even those driven by hedonic or social motivations, showed a lower Sense of Virtual Community to the Linux Kernel. This result  aligns with Cognitive Evaluation Theory \cite{deci1985cognitive} and might be explained by the conflicting identities and divided loyalties that paid contributors have to both their sponsoring firms and the Linux Kernel community \cite{schaarschmidt2018company}. We hypothesize that these contributors would leave the community if there were no payment to compensate for their participation.
 
\textbf{Implications for OSS communities to retain contributors} 

\review{SVC is associated with practices on \textit{exchanging support} \cite{blanchard2011sense,tonteri2011antecedents}, \textit{creating identities and making identifications} \cite{blanchard2011sense}, \textit{producing mutual cognitive and affective trust} amongst members of a community \cite{blanchard2011sense}, and establishing norms and a \textit{``concertive (but not enforced) control''} \cite{gibbs2019investigating}, in which members of the community become responsible for directing their work and monitoring themselves.}
OSS communities can provide not only online interest groups for members, chat rooms, instant messaging, and discussion forums to encourage community involvement \cite{xu2015empirical} but also online tools with shared spaces for contributors to work ``together" on issues to be able to discuss and collaborate on similar interests. Better interactions can strengthen contributors’ Sense of Virtual Community, especially those seeking social relationships. When the information being exchanged surpasses the technical content and includes socio-emotional support, it shows personal relationships among group members, and finally brings feelings of acceptance by members \cite{blanchard2011sense}. OSS communities should foster exchanging support among members to bring a positive impact on developing SVC \cite{tonteri2011antecedents}. The exchanging support includes technical and social support and happens through comments in pull requests and participation in mailing lists (by either reading or posting messages). Communities can manage pull requests and mailing lists to guarantee that members' posts are not being missed \cite{miller2022did} and that the communication adheres to the code of conduct.

\textbf{Implications for OSS communities to attract newcomers.} 
\review{Exchanging information and providing support to other community members are practices associated with positive feelings toward the community, and members' stronger attachment to the community \cite{blanchard2004experienced}.} Community members can encourage newcomers to become more active and move beyond the stage of `lurker,' enticing them to participate in mailing lists \cite{tonteri2011antecedents} and to start making social connections to establish mutual trust, be known by other contributors, and facilitate the development of their Sense of Virtual Community. Conferences and meetups can help hedonic and socially motivated contributors have fun and increase their social capital.


\textbf{Implications for Research.} This study suggests a positive link between Social and Hedonic motivations and a Sense of Virtual Community. Further, the cluster analysis has detected unobserved heterogeneity within our sample, suggesting that there are different subgroups within the community for which different motivations play a more prominent role. Future work could explore how the challenges faced by contributors influence the development of a Sense of Virtual Community and how a Sense of Virtual Community influences the decisions to stay or leave a project. While we included three control variables, future work can consider additional variables, for example, demographic variables such as age. 

Our study focuses on the Linux Kernel community, which is a limitation to the generalizability of this study; we suggest that our findings provide a useful starting point to conduct similar studies across other specific communities or across OSS developers regardless of which community they partake in. When considering other projects, we also suggest that different project governance models might also play a role in SVC. This study has also demonstrated that payment plays a role in SVC and that minorities and marginalized individuals feel less part of the community. 

Finally, our study has focused on the antecedents of a sense of virtual community in OSS but not on the consequences of it, and this could be a further area of focus in future work. Future work can investigate whether SVC is related to contributor satisfaction and whether a reduced SVC leads to contributors' attrition, thus jeopardizing a community's sustainability.

\section{Limitations and Threats to Validity}
\label{sec:threats}


\textit{Construct Validity.}
We adopted and tailored existing measurement instruments for some constructs based on prior literature. Our analysis of the measurement model confirmed that our constructs were internally consistent and scored satisfactorily on convergent and discriminant validity tests.
\ksedit{In this study we have used respondents' country of residence as a proxy for Power Distance as a dimension of culture as defined by Hofstede \cite{hofstede2011dimensionalizing}. While also used in other studies \cite{cortina2017school}, we acknowledge it is an approximation and not a perfect measure. One potential issue is that }
\review{we do not know how \textit{long} respondents have lived in their current country of residence. Another potential issue is that contributors' culture from where they grew up may differ from their current culture. This is why we report the metric as being surrounded by a specific culture instead of having a specific culture.} \ksedit{Measuring culture in a more precise way is an important avenue for future work in general.}

\textit{Internal Validity.}
Our hypotheses propose associations between different constructs rather than causal relationships, as the present study is a cross-sectional sample study \cite{stol2018abc}. We acknowledge the limitation that our respondents comprise contributors who are more likely to have a sense of virtual community as they dedicated their time to answering the questionnaire, suggesting a response bias. While it is clear that contributors motivated by some intrinsic-social reasons tend to experience a sense of virtual community and that power distance and financial compensation can influence those associations, a theoretical model such as ours cannot capture a complete and exhaustive list of factors. 
Other factors can play a role and our results represent a starting point for future studies.


\textit{External Validity.}
The Linux Kernel is a mature project that has attracted contributors for its value over the years, and while studied frequently and sometimes positioned as a `quintessential' open source project, open source projects can vary in many ways. The specific context of the Linux Kernel project may therefore have impacted the results, which are therefore not necessarily  generalizable to other OSS projects. \review{Nevertheless, theory-building is a continuous and iterative process of proposing, testing, and modifying theory \cite{shull2007guide}, in which a single case study is a first step towards constructing theories. In that sense, it is more valuable to interpret these results as a starting point and seek theoretical generalizability rather than statistical generalizability.
Further, given the very important role that the Linux Kernel project plays in the software industry, this project we argue this is an appropriate starting point; further replication studies can validate and extend our theoretical model.} 

Our survey was conducted online and anonymously, but the numbers are aligned with the overall distribution of the Linux Kernel contributors. The Linux Kernel includes contributions of more than 20,000 developers  \cite{LinuxKernel2020}, and are mostly paid to contribute \cite{corbet2012linux,homscheid2015private}. According to previous research, around 10\% of contributors to Linux Kernel identify themselves as women \cite{bitergia2016survey}, and the majority is from the USA, which is aligned with our sample. The responses were sufficiently consistent to find full or partial empirical support for four hypotheses. 

\section{Conclusion}
\label{sec:conclusion}
A Sense of Virtual Community (SVC) helps individuals feel valued in their community, leading to more satisfied, involved, and committed contributors. While research has identified different motivations and challenges for contributors to OSS, it is unclear how a sense of community is created and what factors impact it.

In this paper, we close this gap by developing a theoretical model for sense of virtual community in OSS through a survey of Linux Kernel contributors. 
We found evidence that a subset of intrinsic motivations (social and hedonic motives) are positively associated with SVC; however, other extrinsic factors such as the country's culture and being paid to contribute can lessen SVC among contributors.
Additionally, those with higher English confidence feel a higher sense of belonging in the community, and contributors who identify as part of a gender minority (non-men), tend to feel less of a sense of virtual community. 

Our results also show heterogeneity in our respondents, suggesting that there are different subgroups within the community for whom different motivations play a more prominent role. This suggests that a ``one size fits all'' approach would not work when designing interventions to create an inclusive, welcoming community. Our SVC model can help researchers and community design interventions by highlighting the different factors that interplay in creating a sense of virtual community in OSS for different subgroups of contributors.

\section*{Acknowledgments} 
We are indebted to
Kate Stewart, Vice President of Dependable Embedded Systems at the Linux Foundation, and Shuah Khan, maintainer and contributor of the Linux Kernel, for their invaluable support and assistance in the survey design and distribution. 
We much appreciate the extensive efforts and time they spent in meetings and reviews, and their thoughtful input to the survey design,
and reaching out to the Linux Kernel community members. 
We are deeply grateful to all respondents of this survey. 
The National Science Foundation partially supports this work under Grant Numbers 1900903, 1901031, 2236198, 2235601, and Science Foundation Ireland grants no. 13/RC/2094-P2 to Lero, the SFI Research Centre for Software, and no. 15/SIRG/3293.

\bibliographystyle{IEEEtran}
\bibliography{references}

\end{document}